\documentclass[conference]{IEEEtran}
\usepackage[letterpaper, left=0.625in, right=0.625in, bottom=1in, top=0.75in]{geometry}
\IEEEoverridecommandlockouts
\usepackage{cite}
\usepackage{amsmath,amssymb,amsfonts}
\usepackage[capitalize]{cleveref}
\usepackage{algorithmic}
\usepackage{graphicx}
\usepackage{textcomp}
\usepackage{xcolor}
\usepackage{float}
\usepackage{caption}
\usepackage{subcaption}

\usepackage{siunitx}
\DeclareSIUnit{\nothing}{\relax}

\title{Deep Reinforcement Learning for Combined Coverage and Resource Allocation in UAV-Aided RAN-Slicing}

\author{\IEEEauthorblockN{
Lorenzo Bellone\IEEEauthorrefmark{1},
Boris Galkin\IEEEauthorrefmark{2},
Emiliano Traversi\IEEEauthorrefmark{1},
and Enrico Natalizio\IEEEauthorrefmark{1}}
\IEEEauthorblockA{\IEEEauthorrefmark{1} Autonomous Robotics Research Center - Technology Innovation Institute, UAE}
\IEEEauthorblockA{\IEEEauthorrefmark{2} CONNECT - Trinity College Dublin, Ireland}
lorenzo.bellone@tii.ae,
galkinb@tcd.ie,
emiliano.traversi@tii.ae,
enrico.natalizio@tii.ae
}

\begin{document}
\maketitle

\begin{abstract}
    Network slicing is a well assessed approach enabling virtualization of the mobile core and radio access network (RAN) in the emerging 5th Generation New Radio. Slicing is of paramount importance when dealing with the emerging and diverse vertical applications entailing heterogeneous sets of requirements. 
    
    5G is also envisioning Unmanned Aerial Vehicles (UAVs) to be a key element in the cellular network standard, aiming at their use as aerial base stations and exploiting their flexible and quick deployment to enhance the wireless network performance. 
    
    This work presents a UAV-assisted 5G network, where the aerial base stations (UAV-BS) are empowered with network slicing capabilities aiming at optimizing the Service Level Agreement (SLA) satisfaction ratio of a set of users.
    The users belong to three heterogeneous categories of 5G service type, namely, enhanced mobile broadband (eMBB), ultra-reliable low-latency communication (URLLC), and massive machine-type communication (mMTC). 
    A first application of multi-agent and multi-decision deep reinforcement learning for UAV-BS in a network slicing context is introduced, aiming at the optimization of the SLA satisfaction ratio of users through the joint allocation of radio resources to slices and refinement of the UAV-BSs 2-dimensional trajectories. The performance of the presented strategy have been tested and compared to benchmark heuristics, highlighting a higher percentage of satisfied users (at least 27\% more) in a variety of scenarios. 
\end{abstract}

\section{Introduction} \label{sec:intro}

The 5th Generation (5G) New Radio (NR) is expected to support a large variety of vertical use cases (e.g., autonomous vehicles, industrial Internet of Things), which come with a broader range of requirements with respect to the traditional mobile broadband services.
The NGNM 5G White Paper \cite{5gwhitepaper} groups these vertical use cases into three categories: i) enhanced Mobile Broadband (eMBB), ii) massive Machine Type Communication (mMTC), and iii) Ultra-Reliable Low-latency Communication (URLLC). 
The compliance with the heterogeneous set of requirements demanded by these services is proven hard to achieve in the monolithic architecture implemented in the 4G system \cite{5goverview}. 
A partitioning of the physical network into several virtual networks looks to be the most suitable approach to provide a customized service to each application and to limit the operation expenses. This design is well known as \textit{network slicing}.

The concept of network slicing has been investigated for some time. With the purpose of providing multiple core networks over the same network infrastructure, DECOR and eDECOR \cite{decor} have been already used in legacy LTE networks. 
However, the slicing of Radio Access Network (RAN) was not considered in these works yet.
Since then, the 5G network slicing concept evolved, by providing a better modularization and flexibility of the network functions.
RAN slicing is now of great interest for the literature, given the ability to support resource isolation among different slices, providing a way to allocate radio resources to the connected User Equipments (UEs) based on the slices they belong to. 
Several RAN slicing approaches have been proposed in the past few years. 
Traditional strategies (\hspace{1sp}\cite{RANslicing1}, \cite{orion})  mainly deal with orthogonal resource allocation in time and/or frequency domain, such that the isolation between different services is guaranteed. Their aim in mainly oriented at allocating virtual Resource Blocks (vRBs) to UEs for intra-slice scheduling and to map the allocated vRBs to Physical Resource Blocks (PRBs) at a second stage.
This two level scheduling enables high scalability and dynamicity in the resource management.

Machine Learning (ML) techniques have also been extensively considered, given their capability to find patterns within the huge amount of data that is exchanged in cellular networks. 
In the literature, two main categories of ML strategies for network slicing have been exploited: i) slice admission control, and ii) cross slice resource allocation. 
In the first category, the network operator has to decide whether to accept or refuse slice creation requests coming with a resource requirement and a service level agreement demand. 
This decision has to be made according to the number of resources already allocated and to the monetary value of each slice.
In \cite{sliceadmissioncontrol}, an RL-based slice admission control strategy is proposed. 
Through a value iteration approach, a single agent has to take the acceptance decision, based on the available resources and the monetary value brought by that slice.
If accepted, a fixed allocation of resources will be performed for that slice. 
In the cross slice resource allocation category, instead, the dynamicity of the users is taken into account. The resources are not statically allocated to the slices, but they change over time, according to the dynamic requirements of the users. 
Bega et al. \cite{DeepCog} introduce "DeepCog", a mobile traffic data analytic tool that is tailored at solving a demand forecasting problem. They are able to predict the capacity that each slice will need in future time slots by the design and training of a Deep Neural Network (DNN) that takes as input the current traffic associated to a slice at a given base station. 
Although this work is mainly oriented at fostering the dynamic allocation of resources to the slices, they do not provide a new algorithm to properly schedule the resources given this prediction. 
In \cite{lstm}, instead, the authors rely on a end-to-end resource allocation framework. 
They design a Reinforcement Learning (RL) algorithm based on A2C with an LSTM neural network, which captures the packets exchanged by the slices at each time-slot. 
This input is then converted to an action, based on the radio allocation for each slice for the next allocation window.

These recent works have proven how ML might enable more powerful strategies for resource allocation in network slicing, exploiting the learning of users' behavioral patterns to predict their requirements and properly manage the radio allocation over time. 
These reasons motivate our willingness to exploit similar strategies in an Unmanned Aerial Vehicle (UAV)-assisted next generation cellular network. 

Many works have already proven how UAVs can foster the capabilities of next generation wireless networks through their easy deployability and strong Line-of-Sight communication links, as meticulously described in \cite{Mozaffaritutorial}. However, there is still a limited number of works dealing with UAVs deployment in a RAN slicing context.

A great example is provided by Yang et al. \cite{Yangetal}, formulating a solution for proactive UAV network slicing. In their work, a Lyapunov-based optimization framework is developed in order to maximize the users' achievable data rates as well as the UAVs' fair coverage. 
This optimization technique proved to be effective in a slicing scenario, but it requires a global knowledge of the environment in space and time domain.
A second relevant contribution for UAV-based network slicing is given by \cite{Choetal}, where they propose a resource allocation algorithm for UAV RAN slicing for eMBB and mMTC users co-existence. While they provide a fair solution for the allocation, the authors do not exploit the mobility of UAVs, which has a great impact on the network performance. 

Our work aims at providing an online learning framework, where a set of UAV-BSs will try to maximize the SLA satisfaction ratio of the users deployed in the environment. Unlike the previous mentioned works, this optimization is achieved through a multi-agent and multi-decision approach, trying to exploit the placement of the base stations and the resource allocation for each slice. In more detail, our contribution can be summarized into three fundamental components:
\begin{enumerate}
    \item we propose a joint placement and resource allocation framework in a UAV-aided RAN slicing system, where eMBB, URLLC, and mMTC users are deployed in the same environment;
    \item a multi-decisional and multi-agent reinforcement learning solution is proposed to address these two problems at the same time, exploiting communication between agents;
    \item a fair comparison between our approach and benchmark solutions is presented, showing the benefits of a learning approach in the considered environment.
\end{enumerate}
\section{Methodology} \label{sec:method}
We consider a downlink UAV-BS system in which users have to be properly allocated with resource blocks by the UAV-BS in order to meet their downlink requirements. The set of UAV-BS is denoted as $\mathcal{U} = \{u_1, u_2 \dots\} \in \mathbb{R}^2$, where each UAV $u$ is free to move in a two-dimensional space, while keeping its height fixed.
The radio bandwidth is split into Physical Resource Blocks (PRBs), consisting of 12 consecutive subcarriers 15KHz wide. Each PRB is 1ms long (known as Transmission Time Interval, TTI), thus carrying 14 symbols per subcarrier \cite{LTE}.

A set of users $\mathcal{G} = \{g_1, g_2, g_3 \dots\} \in \mathbb{R}^2$ is spatially distributed following a given distribution. The UAV-BSs will adapt to the placement of the users that might appear in the environment. In our specific scenario, the users are distributed according to multiple bivariate Gaussian distributions.


Each user can belong to one of the three classes of slices, i.e., \textit{eMBB}, \textit{URLLC}, and \textit{mMTC}, each class having different requirements in terms of desired data rate.
In our scenario, each user has probability $p_{\text{eMBB}}$ to belong to the eMBB, $p_{\text{URLLC}}$ to the URLLC, and $p_{\text{mMTC}}$ to the mMTC slices.
The data rate can be theoretically evaluated starting from the Signal to Interference plus Noise Ratio (SINR) between the base station and the associated users, which depends on the specific channel model. For our work, we took inspiration from the Air-to-Ground channel model presented by Galkin et al.  \cite{Channel_Model} (Eq. \ref{channel_model}), where the authors exploit a multi-agent Deep Reinforcement Learning (MADRL) approach within a multi-UAV network scenario, aiming at the energy efficiency optimization through the UAVs trajectories adjustment. The UAVs are supposed to be equipped with down-tilted antennas with a cone-shaped coverage pattern, while the users are provided with an omni-directional antenna. Furthermore, the wireless channel between UAVs and users will always be LoS.
\begin{equation} \label{channel_model}
\scalebox{1.1}{
$\operatorname{SINR}_{i, j}^t=\frac{p c \mu\left(y_j, u_i^t\right)\left(\left(\left\|y_j-u_i^t\right\|\right)^2+\left(h_i^t\right)^2\right)^{-\alpha / 2}}{\sum\limits_{k \in \mathcal{U} \backslash i} p c \mu\left(y_j, u_k^t\right)\left(\left(\left\|y_j-u_k^t\right\|\right)^2+\left(h_k^t\right)^2\right)^{-\alpha / 2}+\sigma^2}$}
\end{equation}
In Eq. \ref{channel_model} we report the Signal-to-Interference-and-Noise-Ratio (SINR) observed by user $i$ with respect to UAV $j$, as illustrated in \cite{Channel_Model}. 
In this equation, $p$ represents the UAV transmit power, $c$ is the near-field pathloss, $\alpha$ is the pathloss exponent, and $\sigma^2$ is the noise power. 
Furthermore, the function $\mu(y_j, u_i^t)$ returns the value of the antenna gain, which depends on the coordinates of user ($y_j$) and UAV ($u_i$), together with the antenna beamwidth.

Once the SINR of a single user has been evaluated, we can i) associate that user with a base station and ii) map the SINR with the modulation and coding scheme (MCS) and code rate, as shown in Table \ref{table1}.
\begin{table}[H]
    \centering
    \begin{tabular}{|c|c|c|}
    \hline
        SINR & Modulation Scheme & Code Rate\\
        \hline
        $<5.2$ & QPSK & 0.5879 \\
        \hline
        $>5.2$ and $<6.1$ & 16QAM & 0.3691 \\
        \hline
        $>6.1$ and $<7.55$ & 16QAM & 0.4785 \\
        \hline
        $>7.55$ and $<10.85$ & 16QAM & 0.6016 \\
        \hline
        $>10.85$ and $<11.55$ & 16QAM & 0.4551 \\
        \hline
        $>11.55$ and $<12.75$ & 64QAM & 0.5537 \\
        \hline
        $>12.75$ and $<14.55$ & 64QAM & 0.6504 \\
        \hline
        $>14.55$ and $<18.15$ & 64QAM & 0.7539 \\
        \hline
        $>18.15$ and $<19.25$ & 64QAM & 0.8525 \\
        \hline
        $>19.25$ & 64QAM & 0.9257 \\
        \hline
    \end{tabular}
    \caption{Mapping between SINR and MCS \hspace{1sp}\cite{sinrtomcs}}
    \label{table1}
\end{table}
The UAV-user association follows a simple strategy: the user is associated with the base station corresponding with the highest value of SINR, as long as this value is greater than a given threshold, which we fixed at 5dB.

Once the user has been associated, the SINR is mapped to the MCS for the transmission between base station and user, and the data rate can be evaluated through Eq. \ref{Eq1}, where $bps$ are the bits per symbol evaluated through the modulation scheme, $c$ is the code rate, $168$ is the multiplication between the number of sub-carriers in each resource block (RB) and the number of symbols per each sub-carriers (12 and 14 respectively), and $r$ are the associated resource blocks for that user (\hspace{1sp}\cite{sinrtomcs}).
\begin{equation}
    dr = 168 \times \frac{bps\times c \times r}{0.001}
    \label{Eq1}
\end{equation}
The bandwidth of the system is 20MHz, thus the scheduler will allocate 100 PRBs every TTI.

In the considered system, each single UAV-BS is in charge of allocating portion of bandwidth to each slice. This allocation results in a different number of PRBs available per users, according to the slice they belong to. Eventually, a round robin scheduler is in charge of associating the PRBs reserved to one slice to the users belonging to that slice.

The goal of our approach is to maximize the number of users with their Service Level Agreement (SLA) satisfied. In the considered scenario, we simulate this metric as being the number of users whose data rate is greater than a given threshold.

The aim is thus twofold: on one hand, we try to allocate portion of the bandwidth to the slices, by taking into account different parameters, such as the requirements of the users and their position. At the same time, the adopted methodology is able to jointly determine the position that the base station will take. Hence, we solve a multi-task decision problem, by addressing simultaneously the resource allocation and placement of each UAV-BS. 

The objective of this problem is to maximize the overall users' SLA satisfaction, $\delta$. To achieve this goal, we formalize an optimization problem for a scenario where two UAV-BSs are deployed, where the following variables must be properly defined:
\begin{itemize}
    \item $u$ is a matrix of dimensions $N \times N$, with N being the number of possible positions for a single base station. This matrix enumerates all the possible placement combination of the two UAVs. $u_{ij}$ is a binary variable that is equal to one, if the first UAV is in position $i$ and the second UAV is in position $j$, with $i,j, \in \{1,\dots,N\}$;
    \item $bw$ is a $3 \times 2$ matrix. It indicates the bandwidth that each base station allocates for \textit{eMBB}, \textit{URLLC}, and \textit{mMTC} users, respectively. $bw_{em, b}$ is a non negative variable that represents the fraction of bandwidth that the UAV $b$ (with $b\in \{1,2\}$ allocates to eMBB users. A analogous definitions holds for $bw_{ur, b}$ and $bw_{mm, b}$;
    \item $\delta$ is a vector of dimensions $|\mathcal{G}|$. $\delta_g$ is a binary variable that is equal to one if the SLA of user $g$ is satisfied.
\end{itemize}
Furthermore, for sake of readability, we define an additional parameter  $G_{\text{conn}}(\text{slice,}b)$ (with $\text{slice} \in \{em, ur, mm\}$) that represents the number of users associated to base station $b$ than belong to a specific slice. This parameter is needed to perform a round robin scheduling after the radio resources have been allocated to a given slice.
Furthermore, we identify with $\text{slice}(g)$ the slice user $g$ belongs to.

The optimization problem can be formally written as:

\begin{equation}\label{optim}
    max_{u,bw}\sum_{g=\mathcal{G}}\delta_g
\end{equation}
$s.t.$
\begin{subequations}
\allowdisplaybreaks
\begin{align}
    & \sum_{b \in |\mathcal{U}|} \sum_{i, j \in \mathcal{U}} u_{i, j} \times bw_{\text{slice}(g), b} \times \text{bps}_g \frac{100}{G_{\text{conn}}(\text{slice}(g),b)} \nonumber \\
    & \geq th_g \times \delta_g \hspace{1cm} \forall g \in \mathcal{G} \\[1em]
    & bw_{em,b} + bw_{ur,b} + bw_{mm,b} = 1 \hspace{1cm} \forall b \in |\mathcal{U}| \\[1em]
    & bw_{em,b} \geq 0; bw_{ur,b} \geq 0; bw_{mm,b} \geq 0 \hspace{1cm} \forall b \in |\mathcal{U}| \\[1em]
    & \sum_{i,j \in \mathcal{U}}u_{i,j} = 1 \\[1em]
    & \delta_g \in \{0, 1\} \hspace{1cm} \forall g \in \mathcal{G} \\[1em] 
    & u_{i,j} \in \{0, 1\} \hspace{1cm} \forall i,j \in \{0, 1, ... , |\mathcal{U}-1|\}
\end{align}\label{constr}
\end{subequations}

Constraint (4a) defines the data rate that is achievable by one user and compares it to the data rate threshold in order to associate the proper $\delta_g$. The element "bps" is the pre-computed function presented in eq. \ref{Eq1}. The achieved bit per symbols are then multiplied by the number of resource blocks allocated to that user.
This number is retrieved by dividing the available resources allocated to the users' slice ($bw_{\text{slice}(g), b} \times 100$) by the number of users belonging to the same slice that are associated to the same base station as the considered user ($G_{\text{conn}}(\text{slice,}b)$). Constraint (4b) specifies that, for each base station, the sum of the portions of allocated bandwidth should be equal to one.

\begin{figure}[h]
    \centering
    \includegraphics[scale=0.45]{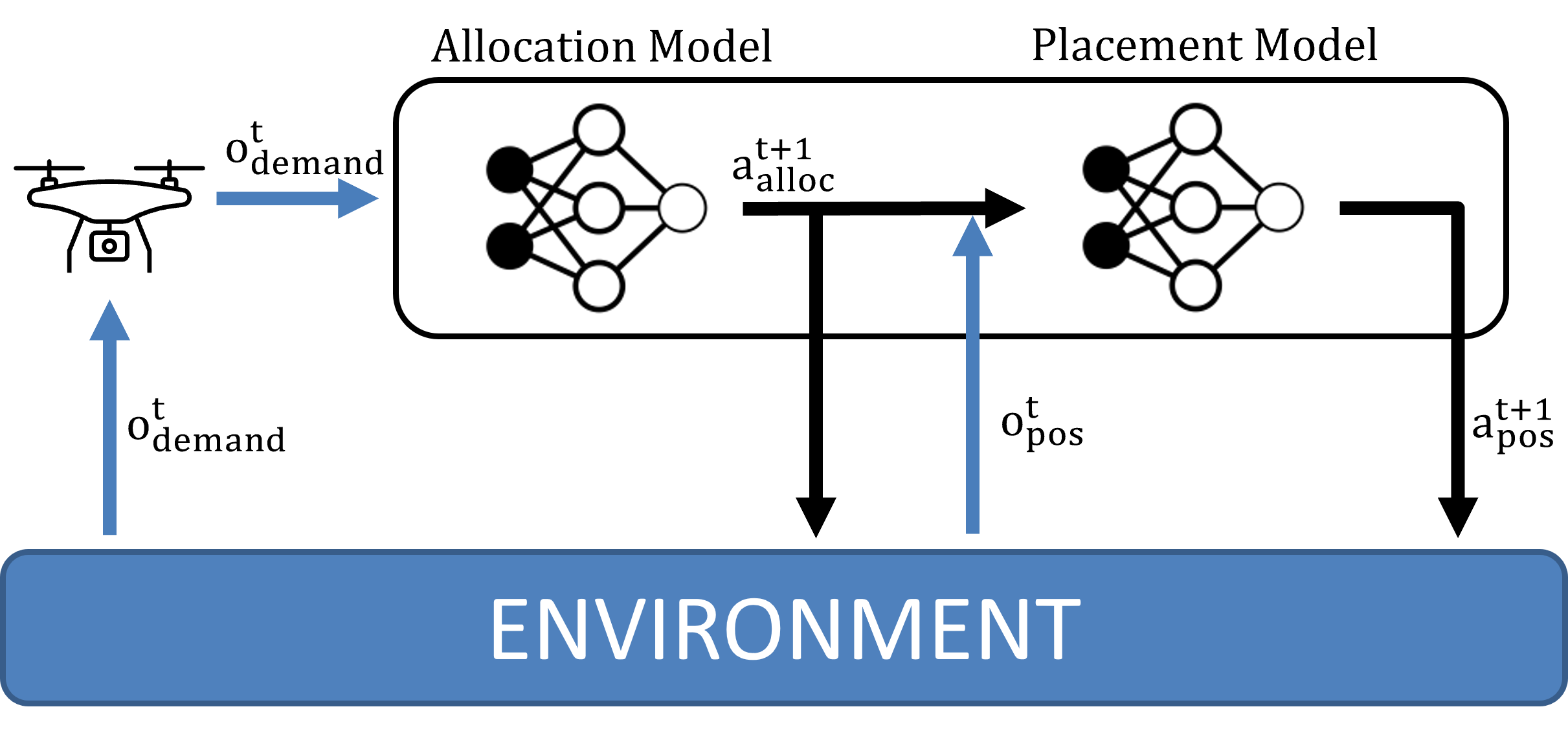}
    \caption{Flowchart showing the information flow in our framework. The UAV-BS takes two observations from the environment. The first one, $O_{demand}^t$, is the input for the resource allocation strategy, while the second one, $O_{pos}^t$, together with the output from the allocation model, provides the input for the placement strategy.}
    \label{fig:flowchart}
\end{figure}

In preliminary tests, we observed that the solution of the presented mathematical model requires a high computation time. As an alternative, we propose a joint placement of UAVs and radio resource allocation carried out by a MADRL approach based on Deep Q-Network (DQN). 
We consider the deployment of physical agents (UAVs), which acquire observations of the surrounding environment and take actions in order to maximize a reward. 
Each physical agent consists of two virtual agents that will follow different policies: the first one is in charge of allocating the radio resources to the associated users, while the second one tries to place the UAV-BS in order to accommodate the best possible number of users. 
An optimal placement of the UAV strictly depends on the maximum number of users that the UAV-BS can satisfy, hence, the two policies are highly correlated. This is the reason why we considered communication between them, by using the decision of radio resources allocation as one of the input for the placement policy. A visual representation of the explained framework can be found in Fig. \ref{fig:flowchart}.

The environment can be modeled as a Markov Decision Process (MDP), where the tuple of state, observation, action and reward is separately defined for the two different policies:
\begin{enumerate}
    \item States: at a certain timestep $t$ the observation from the UAV-BS consists of:
    \begin{itemize}
        \item the aggregated data rate demand from all the associated users from a given slice $s$, $agg\_dem_{s}$;
        \item the distance from all the users $d_{t, g}$;
        \item the relative angle from all the users $\alpha_{t, g}$;
    \end{itemize}
    The first observation is the input of the resource allocation policy, while the other two states, together with the output from the first policy, are the inputs for the placement strategy. 
    Given the multi-agent peculiarity of this scenario, the agents will also get as input the relative position of the other UAV-BS, in order to share information regarding which area of the environment the others are currently covering.
    \item Actions: each UAV-BS decides how to split the available bandwidth between the three slices, according to the users it is covering. This decision can be seen as a choice of three numbers, $\{bw_{eMBB}, bw_{URLLC}, bw_{mMTC}\} \in \mathbb{R}$ such that $bw_{eMBB} + bw_{URLLC} + bw_{mMTC} = 1$. The decision of the bandwidth allocation is part of the input for the placement policy, which will return five possible actions: straight, left, right, back, hover. These actions are turned into discrete movements of a given step-size that the UAV-BS will take at every decision step.
    \item Reward: once each UAV-BS takes the actions regarding resource allocation and placement, they are rewarded according to the level of satisfaction achieved by the associated users. As previously mentioned, in this work we consider the SLA being based on the data rate only. Hence, for every UAV-BS, the reward that is returned at the end of a decision step will be the number of users associated to that base station ($G_{\text{conn}}(b)$) whose SLA is satisfied (eq. \ref{Eq2}), e.g., the number of users whose data rate is higher than a certain threshold $th_g$.
    \begin{equation} \label{Eq2}
        \delta_{b,t} \hspace{0.2cm} = \sum_{g \in G_{\text{conn}}(b)}\mathbb{I}(dr_g>=th_g)
    \end{equation}
\end{enumerate}

In our multi-agent configuration, each UAV-BS has two DQNs, which act as the policies of radio resource allocation and placement. 
Each DQN consists of 3 feed-forward layers with ReLU activation function. 
The inputs of the networks are the states of the described MDP, while their outputs are the actions that the agents take. 
The rewards are used to train the neural network according to the algorithm described in \cite{DQN}. 
Every timestep, each agent collects an observation of the environment and feeds it to its DQN. After the generated action is executed, the agent gets the immediate reward. 
The set of observation, action, reward and next observation is stored in a memory, called replay buffer.
When this dataset is sufficiently populated, it will be periodically sampled in order to train the agent's DQN to maximize the cumulative reward over one episode.
To make the training more robust to noise and speed up the learning convergence, we used a prioritized replay buffer \cite{priobuff}, where the training batches are sampled from the buffer according to an importance weight associated to each of its entries.

\section{Results} \label{sec:results}
To assess the validity of our approach, we performed an extensive simulation campaign through multiple instances of the proposed scenario. 
The number of users and the size of the environment are not fixed, but vary according to the number of UAVs that are deployed. These parameters change in order to preserve the UAVs and users densities (Tab. \ref{tab:envparams}). 
The users are grouped in clusters, whose positions randomly change every new episode.
The number of deployed UAVs is always equal to the number of generated clusters. As the latter, they are randomly placed at the beginning of every new episode. Across the training procedure, the agents will learn how to properly cover the users starting from a random position, and how to fairly allocate radio resources to the slices in order to optimize the SLA satisfaction ratio of the users. Tab. \ref{tab:envparams} provides more details regarding the environment parameters, while information regarding the hyper-parameters of the training procedure can be found in Tab. \ref{tab:hyperparams}.
\begin{table}[h]
    \centering
    \begin{tabular}{|c|c|}
    \hline
    Parameter & Value \\
    \hline
        UAV transmit power $p$ &  1 W\\
        pathloss exponend $\alpha$ & 2.1 \cite{Channel_Model}\\
        UAV half-power beamwidth $\eta$ & 30 deg. \cite{Channel_Model}\\
        Near-field pathloss $c$ & -38.4 dB \cite{Channel_Model}\\
        Noise power $\sigma^2$ & $8 \cdot 10^{-13}$ W\\
        Bandwidth $B$ & 20 MHz\\
        Users density, $|\mathcal{G}|/\text{km}^2$ & 100\\
        UAVs number, $|\mathcal{U}|$ & 1 - 2 - 3 - 4 - 5\\
        UAVs height $h$ & 50 m\\
        UAVs movement step size & 25 m\\
        UAVs density $|\mathcal{U}|/\text{km}^2$ & 8\\
        eMBB users demand $\text{dem}_\text{eMBB}$ & 5 Mbps \cite{demandsvalues}\\
        URLLC users demand $\text{dem}_\text{URLLC}$ & 10 Mbps \cite{demandsvalues}\\
        mMTC users demand $\text{dem}_\text{mMTC}$ & 0.5 Mbps \cite{demandsvalues}\\
        $p_{\text{eMBB}}$, $p_{\text{URLLC}}$, $p_{\text{mMTC}}$ & 20\%, 10\%, 70\% \\
    \hline
    \end{tabular}
    \caption{Environment Parameters}
    \label{tab:envparams}
\end{table}

\begin{table}[h]
    \centering
    \begin{tabular}{|c|c|}
    \hline
    Parameter & Value \\
    \hline
        Episodes per training &  5000\\
        Time-steps per episode & 100\\
        Episodes per testing & 200\\
        Discount factor & 0.95\\
        Learning rate & $10^{-4}$\\
        Initial epsilon value & 1\\
        Epsilon decay value & 0.99995\\
        Minimum epsilon value & 0.01\\
        Replay memory size & 1000\\
        Batch size & 32\\
    \hline
    \end{tabular}
    \caption{Hyper-Parameters for Deep Q-Learning}
    \label{tab:hyperparams}
\end{table}

\begin{figure}[h]
    \centering
    \includegraphics[width=0.47\textwidth]{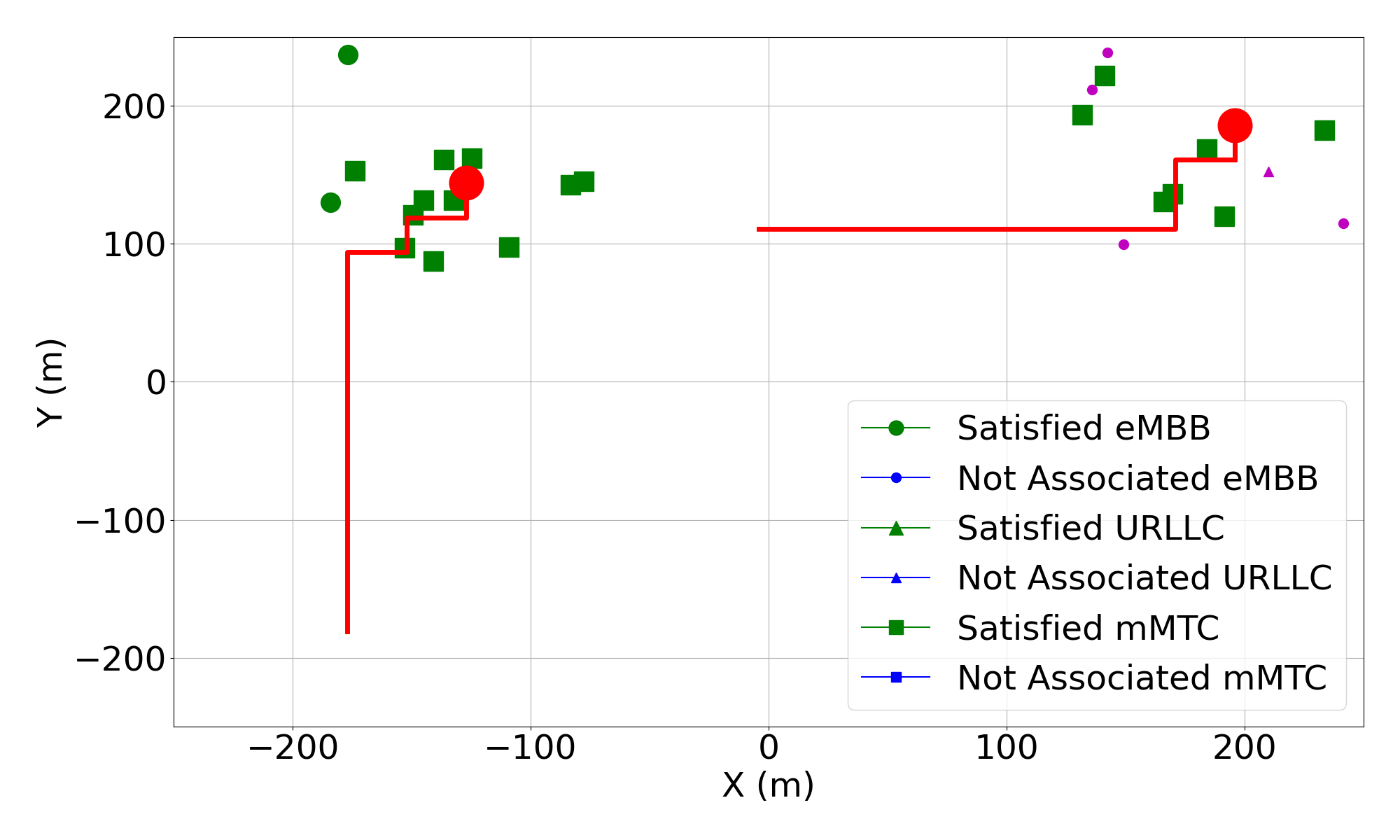}
    \caption{2-dimensional trajectory of 2 UAV-BSs (in red) following their own placement model. The green markers represent the satisfied users after the allocation of resources through our approach. Blue and purple markers respectively represent not associated and not satisfied users.}
    \label{fig:trajectories}
\end{figure}

Fig.\ref{fig:trajectories} shows an example of user coverage with the highlighted trajectory followed by the agents. Furthermore, the green users represent the UEs whose SLA is currently satisfied. 

We decided to compare our approach to three heuristics that have already been used in the literature before (\hspace{1sp}\cite{DeepCog}, \cite{lstm}):
\begin{itemize}
    \item \textbf{Random}. Both UAVs' trajectories and radio resource allocations are randomly selected.
    \item \textbf{R}andom \textbf{A}llocation, \textbf{P}lacement on \textbf{C}entroids (RAPoC). Each UAV will cover one cluster by positioning on its centroid. The radio resource allocation is randomly selected.
    \item \textbf{P}roportional \textbf{A}llocation, \textbf{P}lacement on \textbf{C}entroids (PAPoC). While each UAV will position on one of the centroids, the allocation will be proportional to the aggregated data rate demands of the users that are covered by the flying base station (Eq. \ref{propall}).
\end{itemize}

\begin{equation} \label{propall}
    \text{bw}_{\text{slice}} = \frac{\text{dem}_\text{slice}}{\text{dem}_\text{eMBB} + \text{dem}_\text{URLLC} + \text{dem}_\text{mMTC}}
\end{equation}

\begin{figure}[h]
    \centering
    \includegraphics[width=0.47\textwidth]{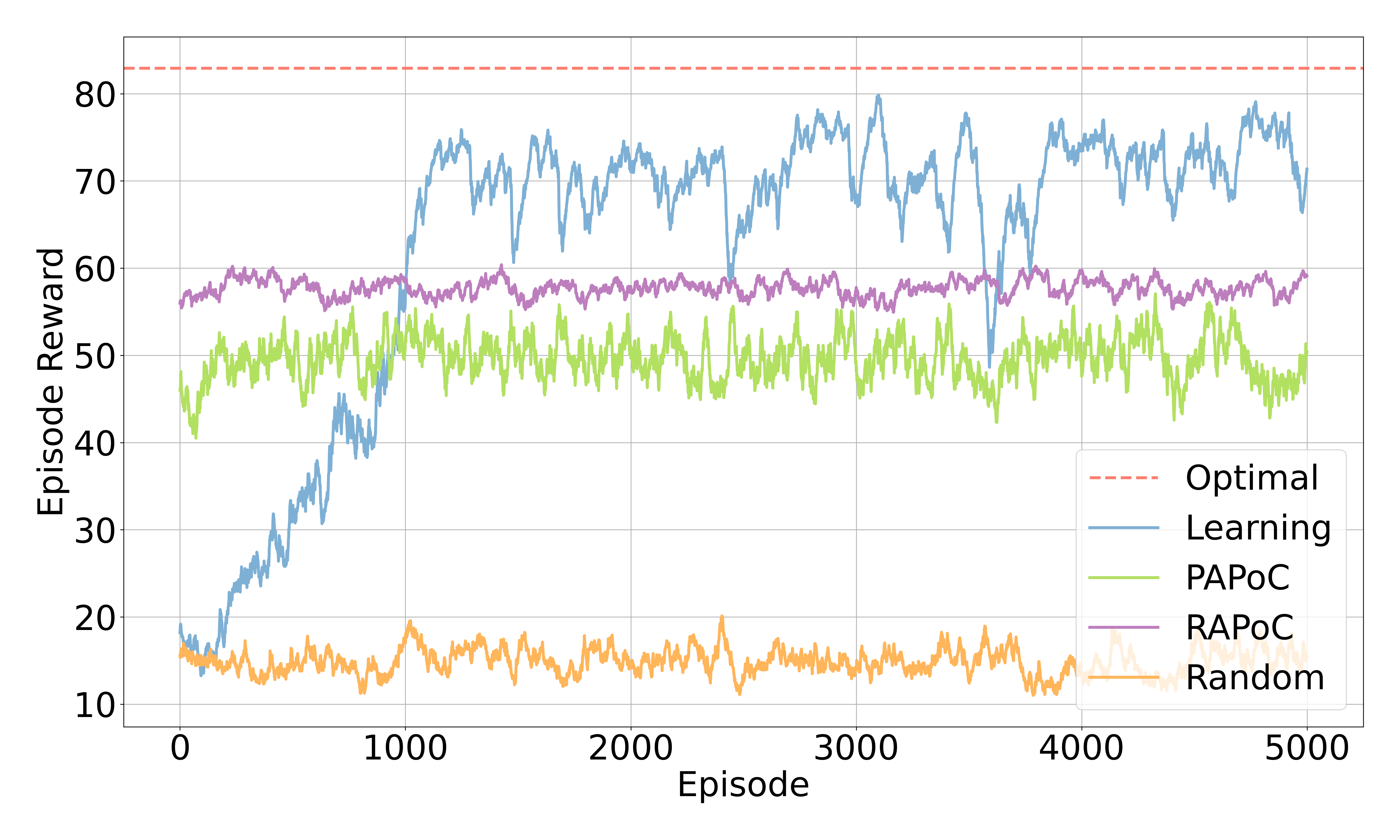}
    \caption{Achieved rewards over episodes, comparing our learning approach (blue line) to three heuristics: random (in yellow), RAPoC (in purple), and PAPoC (in green). The average optimal solution is also depicted in pink.}
    \label{fig:rewardsoverep}
\end{figure}

We have compared our learning approach to the presented heuristics by measuring the average cumulative reward achieved by the whole set of users while exploiting the different policies. 
Fig. \ref{fig:rewardsoverep} shows the rewards achieved through the episodes in the scenario where two clusters of users are deployed. Furthermore, we show the average upper bound in the achieved reward, obtained by deploying a genie-aided solution obtained after solving to optimality the optimization model~\eqref{optim}-\eqref{constr}.

\begin{figure*}[ht!]
     \centering
     \includegraphics[scale=0.2]{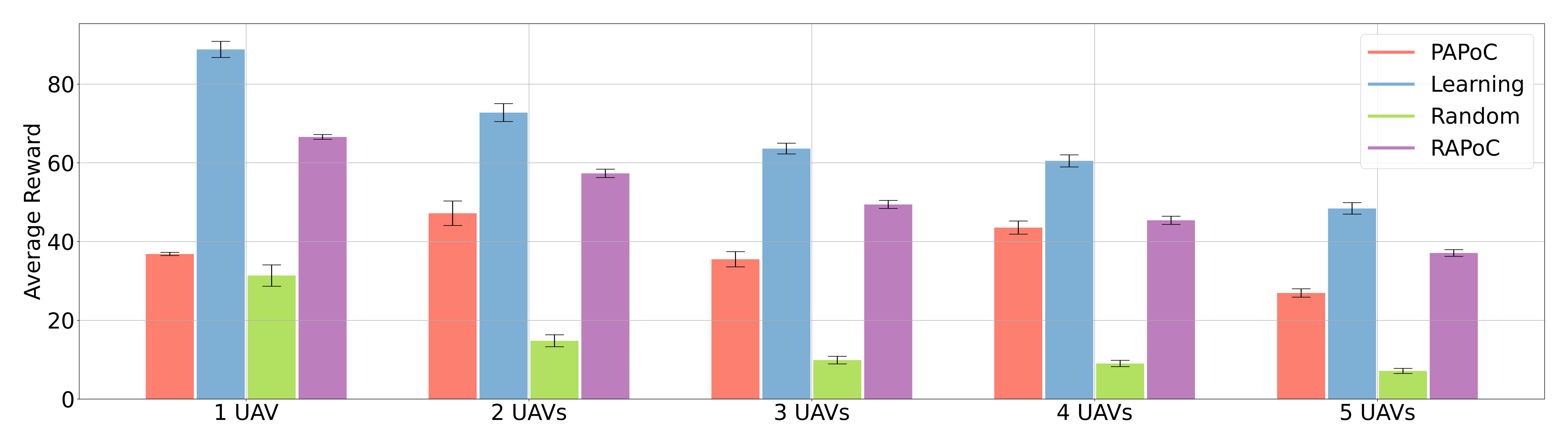}
     \caption{Average reward and 95\% confidence interval achieved with our approach and compared to the other three heuristics by changing the number of UAV-BSs and the number of clusters of users from 1 to 5.}
     \label{fig:barplots}
\end{figure*}

At the beginning of the training process, the learning algorithm (blue line) performs similarly to the random heuristic (in red) because of the initial exploration phase carried out through an $\epsilon$-greedy algorithm. After about 200 episodes, the policy starts to exploit the knowledge acquired, showing an increasing trend in the cumulative rewards that ends up outperforming the other heuristics after about 1000 episodes. 
The fluctuations present in the learned policy are due to the randomization of users' positions at the beginning of every new episode. This proves that, on average, our approach is also able to properly generalize, outperforming the results obtained from the presented heuristics in new environments that have not been observed before.
In order to check the reliability of this algorithm to larger sets of agents, we tested the learning approach on a range of UAVs population spanning from 1 to 5. Figure \ref{fig:barplots} presents the average and the 95\% confidence interval of the cumulative rewards obtained by the agents over 200 episodes, each plot corresponding with a different fleet size. 

From these latest results, it is possible to appreciate the robustness of our strategy with respect to a larger set of agents. The average reward will always outperform the other methods, with the correspondent confidence intervals that never overlap. Even in the most challenging scenario, when 5 UAVs are deployed causing a higher interference in the environment and a larger observation space, the learning approach is able to achieve a reward that is 30.5\% higher than the best result obtained from the heuristics.

\section{Conclusion} \label{sec:conclusion}

In this work, we presented a multi-agent and multi-decision deep RL approach in a UAV-aided 5GNR network slicing scenario, with the aim of optimizing the SLA satisfaction ratio of the deployed users. 
We demonstrated how the learning approach can outperform heuristics that are commonly employed in the literature, by making smarter choices for the placement of UAVs and radio resource allocation for each slice. Furthermore, our learning strategy showed a good generalization capability, adapting the behavior of the agents according to the number of other agents and to different users' distributions. The promising results obtained throughout this research encourage us to investigate deeper into different directions:
\begin{enumerate}
    \item partial observability: the agents do not have a full understanding of the environment's state, but rely on partial observations that can be fostered by communication between base stations;
    \item bigger environment and number of agents/users: a proper study on larger environments is needed to assess the scalability of the system with respect to larger environments and to more dense fleets;
    \item different users' mobility model: in this work we focused on static UEs, while the presence of mobile users would deliver a more realistic and challenging scenario.
\end{enumerate}

\bibliographystyle{IEEEtran}
\bibliography{bibliography}

\end{document}